\documentclass[11pt]{article}

\usepackage[a4paper,margin=2.25cm]{geometry}
\usepackage[T1]{fontenc}
\usepackage{setspace}
\usepackage[switch]{lineno} 

\newif\ifreviewversion
\reviewversionfalse

\usepackage{newtxtext,newtxmath}
\usepackage{microtype}
\usepackage{graphicx}
\usepackage{booktabs}
\usepackage{tabularx}
\usepackage{longtable}
\usepackage{multirow}
\usepackage{array}
\usepackage{amsmath}
\usepackage[super,sort&compress]{natbib}

\usepackage{xcolor}
\definecolor{NatureBlue}{RGB}{0,45,180}

\usepackage[
  colorlinks=true,
  linkcolor=NatureBlue,
  citecolor=NatureBlue,
  urlcolor=NatureBlue,
  breaklinks=true
]{hyperref}

\usepackage[nameinlink,noabbrev]{cleveref}
\usepackage{caption}
\usepackage{float}
\usepackage{placeins}
\usepackage{enumitem}

\newcolumntype{P}[1]{>{\raggedright\arraybackslash}p{#1}}
\newcolumntype{Y}{>{\raggedright\arraybackslash}X}

\newcommand{\SupFig}[1]{(Supplementary Fig.~#1)}
\newcommand{\SupTable}[1]{(Supplementary Table~#1)}
\newcommand{\SupNote}[1]{(Supplementary Note~#1)}

\newcommand{\csPCa}{clinically significant prostate cancer}

\newcommand{\RenjiHospital}{Ren Ji Hospital, Shanghai Jiao Tong University School of Medicine}
\newcommand{\ChanghaiHospital}{Changhai Hospital, Naval Medical University}
\newcommand{\ShanghaiSecondHospital}{Shanghai Second People's Hospital}
\newcommand{\HangzhouBayHospital}{Ningbo Hangzhou Bay Hospital}

\newcommand{\CodeReviewURL}{https://pcnkfl2i3xls.feishu.cn/file/EQysbKKYDoWgZdxhhbMcbjsqnib}
\newcommand{\RenjiIRB}{KY2018-212-s2}

\graphicspath{{Figures/}}

\begin{document}

\ifreviewversion
  \doublespacing
  \linenumbers
\else
  \setstretch{1.08}
\fi

\begin{center}

{\LARGE\bfseries
A cross-modal generative model for incomplete and degraded prostate MRI with multicentre clinical validation
\par}

\vspace{0.8em}

{\large
Siyuan Ma$^{1,\dagger}$,
Liang He$^{2,\dagger}$,
Mengying Zhu$^{3,\dagger}$,
Yi Chai$^{4,\dagger}$,
Mengyao Lyu$^{1,\dagger}$,
Haowei Wang$^{5,\dagger}$,
Qizhen Lan$^{6,\dagger}$,
HaoBo Sun$^{7}$,
Qixin Zhang$^{1}$,
Jingli Chen$^{3}$,
Xiaobing Wei$^{3}$,
Jiaming Liu$^{8}$,
Guiqin Liu$^{3}$,
Qianwen Zhang$^{9}$,
Yang Liu$^{1,*}$,
Dacheng Tao$^{1,*}$,
Guangyu Wu$^{3,*}$
\par}

\vspace{0.7em}

{\small

$^{1}$College of Computing and Data Science, Nanyang Technological University, Singapore\\

$^{2}$School of Computer Science and Technology, Tongji University, Shanghai, China\\

$^{3}$Department of Radiology, Ren Ji Hospital, Shanghai Jiao Tong University School of Medicine, Shanghai, China\\

$^{4}$National University of Singapore, Singapore\\

$^{5}$Department of Medical Oncology, Shanghai East Hospital, Tongji University School of Medicine, Shanghai, China\\

$^{6}$McWilliams School of Biomedical Informatics, The University of Texas Health Science Center at Houston (UTHealth Houston), Houston, Texas, USA\\

$^{7}$Jiangsu Key Laboratory of Urban ITS, Jiangsu Province Collaborative Innovation Center of Modern Urban Traffic Technologies, School of Transportation, Southeast University, Nanjing 210096, China\\

$^{8}$Department of Radiology, Ningbo Hangzhou Bay Hospital, Ningbo, Zhejiang, China\\

$^{9}$Department of Radiology, Changhai Hospital, Naval Medical University, Shanghai, China\\[0.5em]

$^{\dagger}$These authors contributed equally to this work.\\[0.25em]

$^{*}$Co-corresponding authors: Yang Liu, Dacheng Tao and Guangyu Wu.\\
Correspondence e-mail: \href{mailto:danielrau@163.com}{danielrau@163.com} (G.W.)

}

\end{center}
\vspace{0.7em}

\begin{abstract}
Missing or degraded sequences can limit prostate multiparametric MRI. We developed MSCNet, a sequence-conditioned cross-modal generative framework for reconstructing unavailable contrasts and restoring degraded acquisitions. Across ten completion tasks, task-specific MSCNet achieved mean structural similarity of 0.818 versus 0.798 for the strongest task-matched comparators; matched-capacity analyses showed larger differences in lesion fidelity and boundary preservation. In a blinded 1,000-case reader study, overall image quality met the prespecified non-inferiority criterion for DWI, ADC and T2W completion, but not T1W. In a separate 200-case diagnostic assessment, AUCs for clinically significant cancer were 0.860 with acquired images, 0.841 with MSCNet and 0.797 with baseline-generated images. A locked 186-case three-hospital cohort supported multicentre transportability. These retrospective results support quality-controlled cross-modal reconstruction as an adjunct to acquired prostate MRI.
\end{abstract}

Multiparametric magnetic resonance imaging (mpMRI) has changed the diagnostic pathway for suspected prostate cancer by improving detection of clinically significant disease and reducing unnecessary biopsy.\cite{1,2,3,4} Its value arises from the joint interpretation of sequence-specific evidence. T2-weighted imaging (T2W) depicts zonal anatomy, gland margins and tumour morphology, whereas diffusion-weighted imaging (DWI) and apparent-diffusion-coefficient (ADC) maps characterize water mobility and cellularity.\cite{1,5} T1-weighted imaging (T1W) provides complementary information on haemorrhage and pelvic background. The Prostate Imaging Reporting and Data System (PI-RADS) formalizes this sequence-dependent interpretation.\cite{1} The diagnostic integrity of an examination therefore depends not only on the presence of disease, but also on whether the required contrasts are complete, spatially aligned and interpretable.

Sequence quality remains variable in routine practice. DWI is particularly vulnerable to susceptibility-induced distortion, signal dropout, motion and ghosting, and these effects can become more conspicuous at higher field strengths despite gains in signal-to-noise ratio.\cite{5,6} Motion, patient preparation, scanner hardware and protocol heterogeneity can also degrade T2W or make an otherwise complete examination locally non-diagnostic. PI-QUAL and PI-QUAL version 2 were developed because image quality directly affects lesion detection, confidence in a negative examination and downstream biopsy planning.\cite{7,8} In referral workflows, where protocols and sequence availability vary, an incomplete or degraded contrast can therefore create a practical choice between interpreting an uncertain examination and repeating the scan.

Cross-modal synthesis offers a route to estimate unavailable information from the remaining sequences. Earlier work established modality-invariant latent representations, hetero-modal learning, multi-input adversarial synthesis and hybrid or unified fusion strategies for missing-modality imputation.\cite{11,12,13,14,15} Generic adversarial and image-translation models can generate visually plausible outputs,\cite{18,19,37} but perceptual sharpness alone does not establish anatomical or diagnostic fidelity. Most previous evaluations have focused on brain MRI, fixed input configurations or image-level similarity. Prostate MRI poses a stricter problem: a clinically useful system must handle different available-sequence combinations, preserve target-specific anatomy and lesion contrast, remain stable under acquisition artefacts, expose uncertainty and be assessed against radiologist and diagnostic endpoints.

We developed MSCNet as a sequence-conditioned reconstruction framework that preserves modality-specific representations, fuses available contrasts through target-conditioned gates and combines global-context modelling with attention-guided decoding and edge-aware supervision. Evaluation separated physically coupled DWI--ADC tasks from cross-contrast synthesis and used capacity- and objective-matched controls, modern generative baselines, a shared-routing variant, physics-aware controls and retrieval. Clinical relevance was assessed through anatomical and lesion-level endpoints, same-patient repeat scans, a blinded reader study, cancer discrimination, a locked three-hospital external cohort, failure adjudication and selective prediction.

\section*{Results}

\subsection*{Study design and a cross-modal reconstruction framework}

\begin{figure}[!htbp]
\centering
\includegraphics[width=\textwidth,height=0.72\textheight,keepaspectratio]{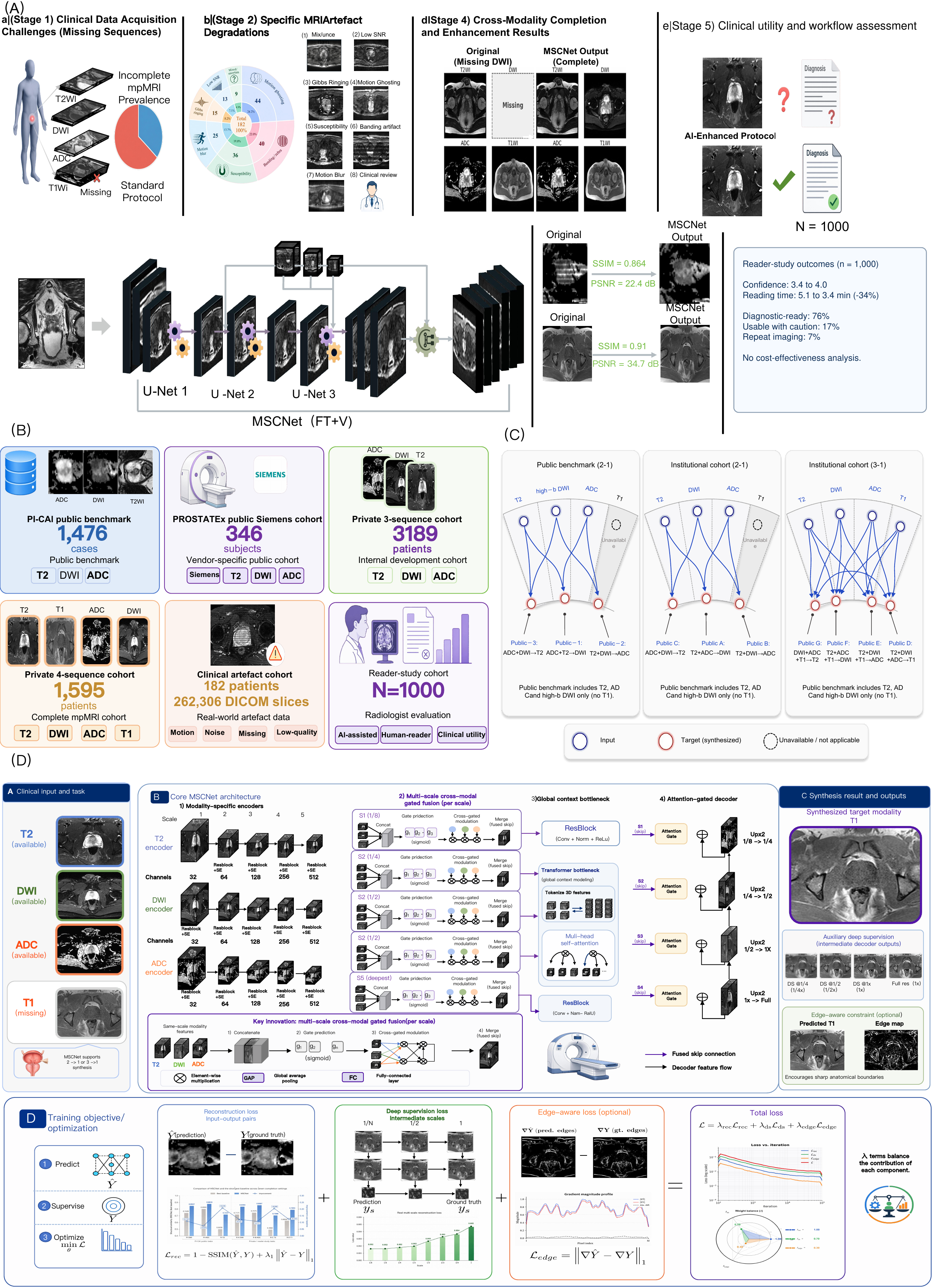}
\caption{\textbf{Study overview, cohorts, completion tasks and MSCNet framework.}
\textbf{A,} Clinical motivation and workflow for incomplete or degraded prostate mpMRI, cross-modal completion or enhancement, and radiologist review.
\textbf{B,} Public, institutional, artefact and reader-study cohorts.
\textbf{C,} Matrix of 2-to-1 and 3-to-1 completion tasks.
\textbf{D,} MSCNet architecture, synthesis outputs and optimization objectives, including modality-specific encoders, multi-scale cross-modal gated fusion, global-context modelling, attention-gated decoding, deep supervision and edge-aware reconstruction.}
\label{fig:study}
\end{figure}

The study addressed two related failure modes of prostate mpMRI (Fig.~\ref{fig:study}). In multi-sequence completion, one target sequence was treated as unavailable and reconstructed from the remaining sequences. In virtual enhancement, an acquired sequence affected by artefact or global degradation was restored using complementary information from the other contrasts. The public and institutional cohorts supported three 2-to-1 tasks in T2W, DWI and ADC, and four 3-to-1 tasks after inclusion of T1W. The principal analysis trained one model for each target configuration; a secondary shared model used a target token and modality mask to process all ten configurations with one set of weights. A clinical artefact cohort, 204 same-patient repeat acquisitions, a 1,000-case reader study, a 200-case diagnostic assessment and a pooled 186-case three-hospital external cohort (site A, 72 cases; site B, 64; site C, 50) extended evaluation beyond image similarity (Extended Data Table~\ref{ext:cohorts}).

MSCNet was designed around the observation that MRI contrasts are not interchangeable image channels. Separate encoders first retained modality-specific representations. At each feature scale, target-conditioned gates then controlled how much information from each available sequence entered the fused skip representation. A Transformer bottleneck modelled long-range context,\cite{21} and an attention-gated decoder\cite{22} reconstructed the target while suppressing features that were irrelevant or dominated by artefact. Squeeze-and-excitation blocks\cite{23} and deep supervision\cite{24} supported channel recalibration and stable multiscale optimization. Ablation analysis showed that the full model outperformed removal or replacement of each component across distortion, structural and perceptual metrics; the largest aggregate degradation followed removal of the global-context bottleneck, while gated fusion and the edge-aware branch contributed complementary gains \SupFig{2}. The ablation pattern assigned complementary roles to global anatomical context, target-conditioned cross-modal routing and boundary-sensitive supervision.

Data governance was treated as part of the modelling problem. Splits were defined at the patient level, and sequence pairs underwent orientation checks, anatomical matching and deformable correction before inclusion. Registration correction reduced landmark errors across sequence pairs and anatomical levels, whereas ADC-DWI pairs remained the most intrinsically aligned. The quality-control pipeline excluded 272 cases, most commonly for missing sequences, severe artefact or sequence mismatch \SupFig{4}; detailed cohort governance and exclusion criteria are reported in \SupTable{1}. PROSTATEx served as a vendor-specific public-cohort analysis and was excluded from the independent external-validation claim because of its relationship to PI-CAI.

\FloatBarrier

\subsection*{Reconstruction gains occurred in anatomical and lesion-relevant regions}

\begin{figure}[!htbp]
\centering
\includegraphics[width=\textwidth,height=0.72\textheight,keepaspectratio]{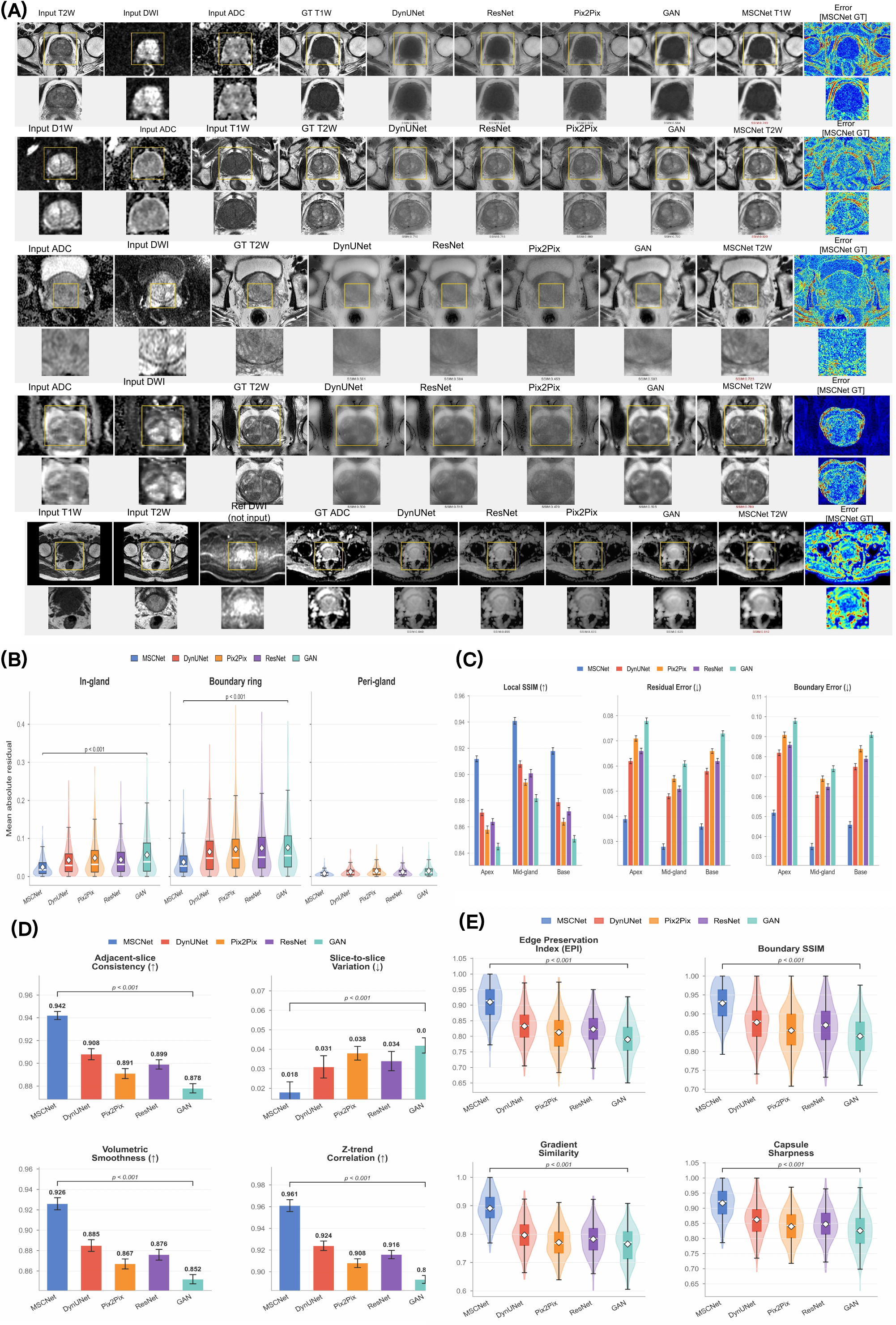}
\caption{\textbf{Visual and spatial behaviour of multi-sequence prostate MRI completion.}
\textbf{A,} Input sequences, acquired reference targets, baseline outputs, MSCNet outputs, residual maps and enlarged regions.
\textbf{B,} Residual decomposition in the gland, boundary ring and peri-gland regions.
\textbf{C,} Local structural similarity, residual error and boundary error at the apex, mid-gland and base.
\textbf{D,} Adjacent-slice consistency, slice-to-slice variation, volumetric smoothness and z-trend correlation.
\textbf{E,} Edge preservation, boundary SSIM, gradient similarity and capsule sharpness.}
\label{fig:visual}
\end{figure}

Representative cases showed distinct failure patterns among the comparison methods (Fig.~\ref{fig:visual}A). Regression baselines produced smooth gland interiors and softened capsule margins; adversarial baselines recovered sharper texture but introduced local signal instability. MSCNet reduced both behaviours. The residual maps were less spatially extensive, and magnified regions retained gland contours and intraprostatic structure across T1W, T2W, DWI and ADC targets.

The regional analyses established that these gains were not driven by uniform background. Absolute residuals were lowest for MSCNet within the gland and boundary ring, where errors are most likely to affect lesion conspicuity or zonal assessment (Fig.~\ref{fig:visual}B). The advantage persisted at the apex, mid-gland and base (Fig.~\ref{fig:visual}C). In the local volumetric analysis, adjacent-slice consistency was 0.942 for MSCNet, compared with 0.908 for DynUNet and 0.874 for the GAN baseline; slice-to-slice variation was correspondingly lower (0.018 versus 0.031 and 0.050). Volumetric smoothness and z-trend correlation showed the same ordering (Fig.~\ref{fig:visual}D). Boundary-specific measures, including edge preservation, boundary SSIM, gradient similarity and capsule sharpness, were also highest for MSCNet (Fig.~\ref{fig:visual}E). Together, these results indicate that improved similarity reflected preservation of coherent anatomy rather than only denoising of large low-information regions.

Lesion-centred analyses provided a second test of anatomical fidelity. Correlations between acquired and generated lesion contrast were 0.959 for contrast-to-noise ratio, 0.962 for lesion-region signal-to-noise ratio and 0.966 for lesion-to-background contrast ratio, compared with 0.792, 0.821 and 0.853 for Pix2Pix \SupFig{5}. Line profiles showed that MSCNet followed acquired lesion peaks, troughs and transition zones more closely than the comparison methods. Radiologist-rated lesion conspicuity was 4.29 for MSCNet and 4.47 for acquired reference images, compared with 3.62 for the diffusion baseline, 3.33 for the residual-transformer baseline and 2.73 for Pix2Pix \SupFig{5}. The remaining gap from acquired images was concentrated in small lesions, uncertain boundaries and difficult T2W reconstruction, which became important in the subsequent safety analysis.

\FloatBarrier

\subsection*{Performance across targets, cohorts and acquisition strata}

\begin{figure}[!htbp]
\centering
\includegraphics[width=\textwidth,height=0.72\textheight,keepaspectratio]{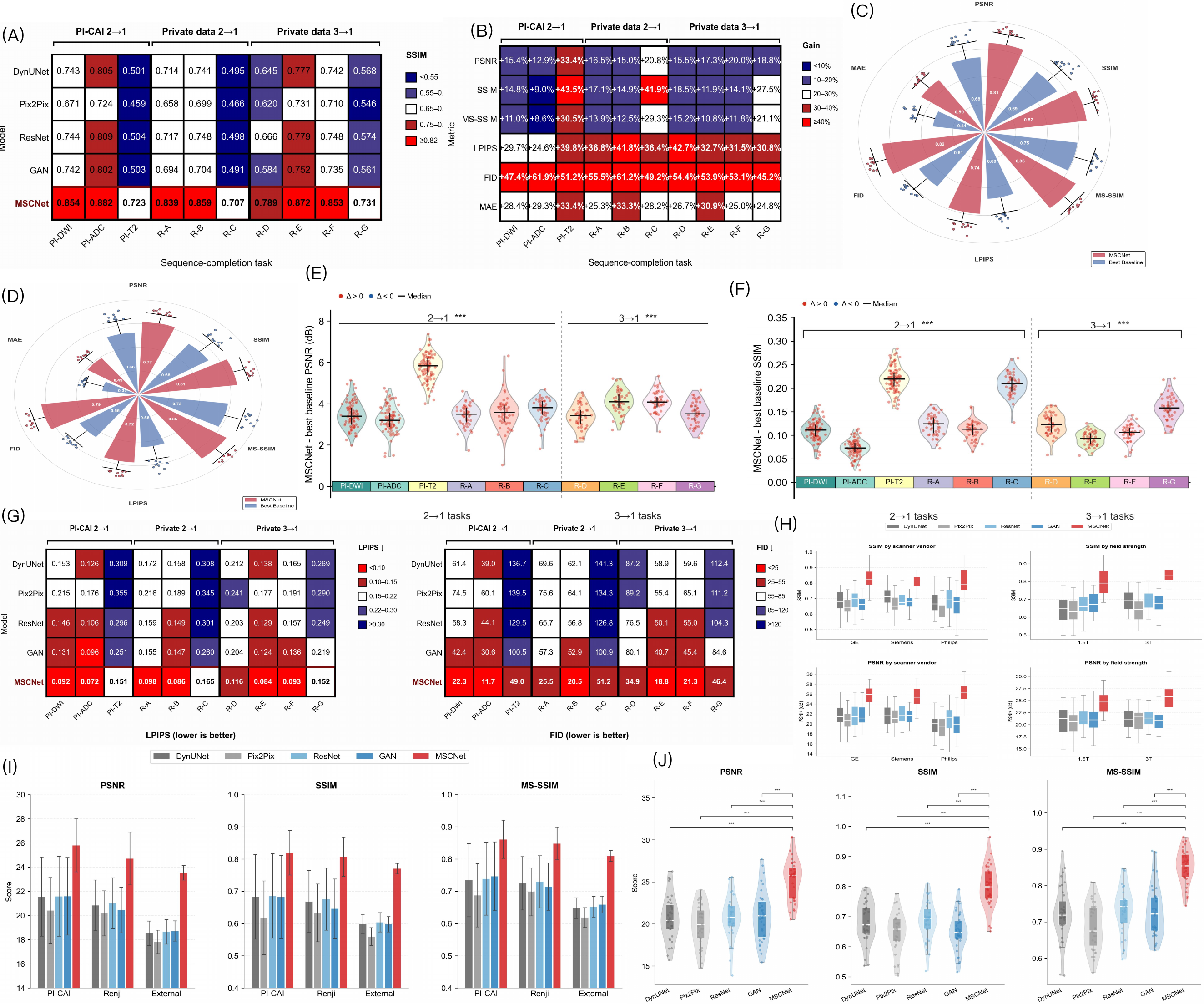}
\caption{\textbf{Optimized quantitative performance and clinical-anatomical effect size.}
\textbf{A,} Task-level SSIM for MSCNet and the strongest task-specific non-MSCNet comparator.
\textbf{B,} Absolute SSIM gain for each completion task.
\textbf{C,} Grouped macro SSIM for physical-coupling, cross-contrast and T1W tasks.
\textbf{D,} Task-level LPIPS for MSCNet and the strongest task-specific comparator.
\textbf{E,} Matched-capacity analysis on the locked subset, showing whole-image SSIM, lesion fidelity and boundary SSIM.
\textbf{F,} Internal and external-hospital SSIM, including the three external sites and pooled estimate.}
\label{fig:quantitative}
\end{figure}

MSCNet achieved the highest SSIM in all ten completion tasks after expansion of the comparison set to unified synthesis, parameter-matched Transformer, masked-diffusion, structure-aware latent-diffusion, loss-matched DynUNet and retrieval or signal-model controls (Fig.~\ref{fig:quantitative}A--C and Extended Data Table~\ref{ext:quantitative}). Because ADC and DWI share diffusion-derived information, we prespecified three task groups rather than interpreting every task as equally difficult. Mean SSIM was 0.867 for physically coupled DWI--ADC tasks, 0.727 for cross-contrast T2W tasks and 0.796 for T1W completion. Relative to the strongest non-MSCNet model selected within each task, the corresponding gains were 0.018, 0.024 and 0.020, respectively \SupTable{14}. The ten-task macro-average was 0.818 versus 0.798 for the strongest task-wise comparators, an absolute gain of 0.020. The largest group-level advantage occurred in the cross-contrast tasks, where the target could not be recovered from a direct diffusion signal relationship.

Matched controls separated the contribution of model capacity, optimization and sequence-conditioned routing. Relative to the parameter-matched Transformer on the locked subset, MSCNet increased lesion fidelity from 0.87 to 0.92 and boundary SSIM from 0.853 to 0.908 while reducing LPIPS from 0.108 to 0.086 (Fig.~\ref{fig:quantitative}E and \SupTable{11}). These localized differences provide a clinical-anatomical anchor for the smaller whole-image SSIM effect without treating SSIM as a diagnostic surrogate. Prespecified module ablation yielded aggregate SSIM values of 0.931 for the full model, 0.917 without the edge-aware branch, 0.912 without gated fusion, 0.906 with direct concatenation and 0.899 without the Transformer bottleneck \SupTable{11}. Target-specific optimized signal-model and retrieval controls also remained below MSCNet on their applicable held-out tasks \SupTable{15}.

A secondary shared variant tested unified input--target routing with one set of weights. MSCNet-Shared achieved mean SSIM 0.804, compared with 0.818 for optimized task-specific MSCNet, while reducing storage from ten 154.1-million-parameter models to one 158.3-million-parameter model \SupTable{12}. Unified pretraining on 3,750 development cases followed by task-specific fine-tuning reached SSIM 0.818 and LPIPS 0.108, whereas few-shot, frozen-encoder and unseen-combination performance remained lower \SupTable{13}. The shared model therefore provides a unified routing extension across evaluated configurations rather than evidence of unrestricted zero-shot behaviour.

Generality was examined across public, institutional, temporal and hospital-level settings (Fig.~\ref{fig:quantitative}F and \SupFig{3}). Correlations across scanner and protocol strata ranged from 0.86 to 0.94. In the pooled three-hospital external cohort, the macro-average across tasks A--C was SSIM 0.791, LPIPS 0.124 and MAE 0.061, compared with 0.809, 0.113 and 0.056 in the internal held-out cohort \SupTable{16}. This shift supports transportability of the reconstruction signal while leaving prospective workflow validation and hospital-specific monitoring necessary.

\FloatBarrier

\subsection*{Clinical artefact transfer extended the model from completion to enhancement}

\begin{figure}[!htbp]
\centering
\includegraphics[width=\textwidth,height=0.72\textheight,keepaspectratio]{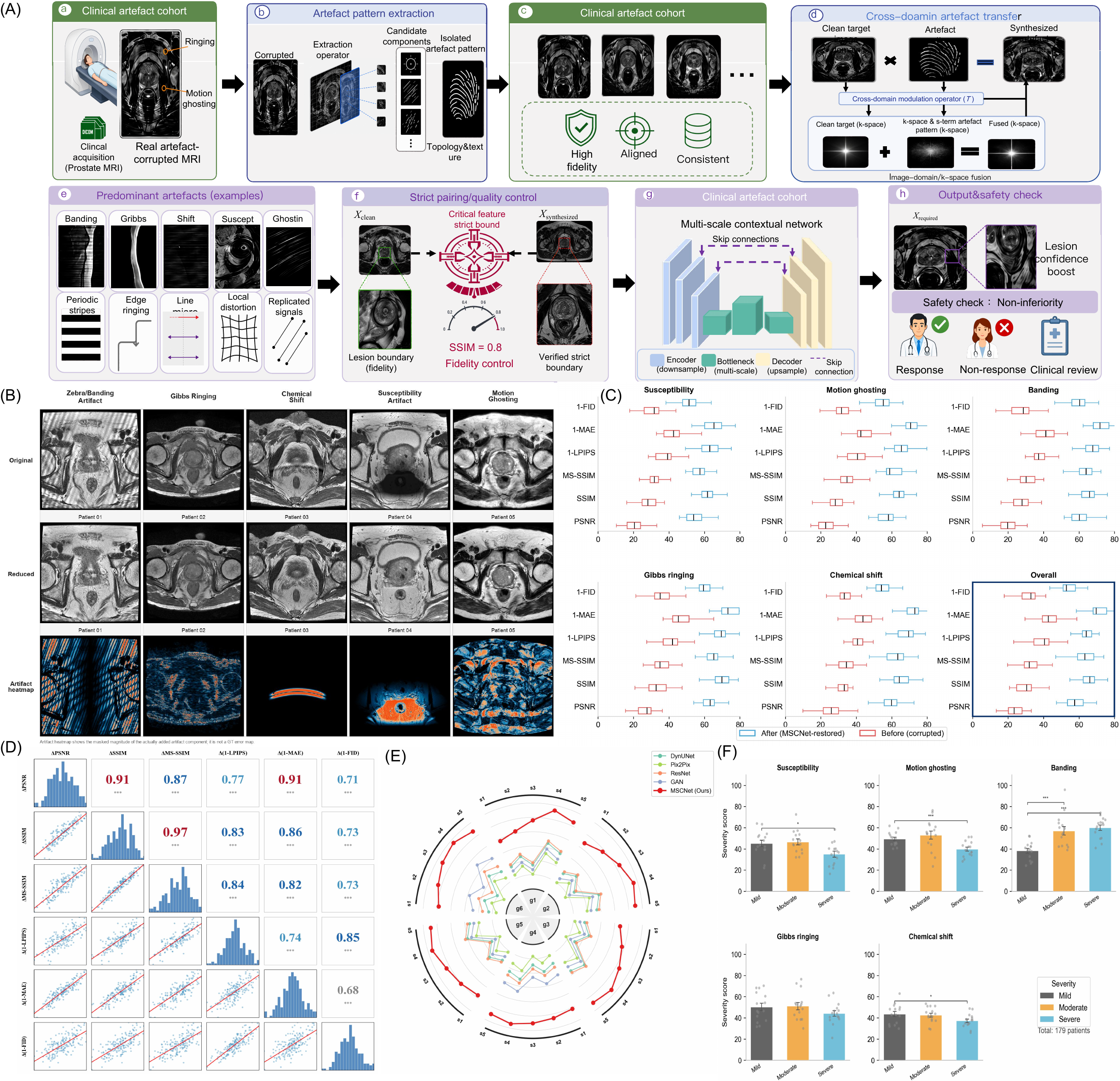}
\caption{\textbf{Clinical artefact transfer and virtual enhancement.}
\textbf{A,} Workflow from clinical artefact collection and pattern extraction to cross-domain transfer, strict pairing, restoration and safety review.
\textbf{B,} Representative susceptibility, motion, banding, Gibbs-ringing and chemical-shift cases before and after restoration.
\textbf{C,} Before-and-after image-quality distributions by artefact type.
\textbf{D,} Correlation among distortion, structural and perceptual metrics.
\textbf{E,} Comparison with baseline restoration models.
\textbf{F,} Artefact-type and severity distribution in the clinical cohort.}
\label{fig:artefact}
\end{figure}

Missing sequences are only one manifestation of incomplete information. In practice, a sequence may be present but locally unreliable. We therefore extracted clinically observed artefact patterns from the institutional archive and transferred them to quality-controlled targets under anatomical and fidelity constraints (Fig.~\ref{fig:artefact}A). The procedure covered susceptibility artefact, motion ghosting, banding, Gibbs ringing and chemical shift, rather than only generic noise or blur. A strict pairing step used lesion-boundary checks and an SSIM threshold of 0.8 to reject implausible degraded-reference pairs.

Enhancement improved every artefact category (Fig.~\ref{fig:artefact}B,C). Across the cohort, PSNR increased from 22.7 to 31.6, SSIM from 0.338 to 0.654 and MS-SSIM from 0.368 to 0.684. The higher-is-better transformations of LPIPS, MAE and FID increased from 0.464, 0.414 and 0.316 to 0.736, 0.670 and 0.628, respectively. PSNR gains were observed for susceptibility artefact (21.4 to 29.9), motion ghosting (22.6 to 30.5), banding (19.8 to 31.7), Gibbs ringing (25.2 to 33.3) and chemical shift (24.5 to 32.6). Concordance among the six quality measures and consistent separation from the comparison models indicated that the effect was not attributable to a single metric (Fig.~\ref{fig:artefact}D,E and \SupTable{3}).

The safety analysis changed the interpretation of these improvements. Among radiologist-adjudicated outputs, any safety event occurred in 9.4\% of MSCNet outputs and 28.7\% of baseline outputs. Suspected false lesions occurred in 1.4\% versus 5.8\%, reduced lesion conspicuity in 3.6\% versus 11.2\%, blurred lesion margins in 5.1\% versus 15.6\%, partial lesion erasure in 2.3\% versus 8.4\%, and non-diagnostic outputs in 3.2\% versus 9.3\% \SupFig{1}. Severe motion and susceptibility artefacts were predominantly acquisition-driven, whereas very small lesions were predominantly lesion-driven and sequence misregistration was predominantly registration/model-driven. Thus, enhancement reduced risk but did not remove it; the residual events were structured and potentially detectable. The adjudicated event rates and their operational definitions are summarized in \SupTable{7}.

A separate same-patient repeat-scan analysis tested whether the controlled-transfer result extended to genuine degraded acquisitions. The 182-patient artefact cohort yielded 204 paired acquisitions because one patient could contribute more than one adjudicated degradation stratum. Mean SSIM increased from 0.714 for the degraded acquisition to 0.883 after MSCNet restoration when compared with a clean repeat scan, an absolute increase of 0.169 ($P<0.001$ with patient clustering). Improvements were consistent across motion blur, susceptibility distortion, low-SNR high-b-value DWI, geometric distortion, zipper artefact, ghosting and radiofrequency inhomogeneity \SupTable{17}. This analysis complements rather than replaces the controlled artefact-transfer experiment: the repeat scans provide a clinically acquired reference, whereas the controlled pairs isolate specific corruption classes.

\FloatBarrier

\subsection*{Radiologist assessment separated visual plausibility from clinical usability}

\begin{figure}[!htbp]
\centering
\includegraphics[width=\textwidth,height=0.72\textheight,keepaspectratio]{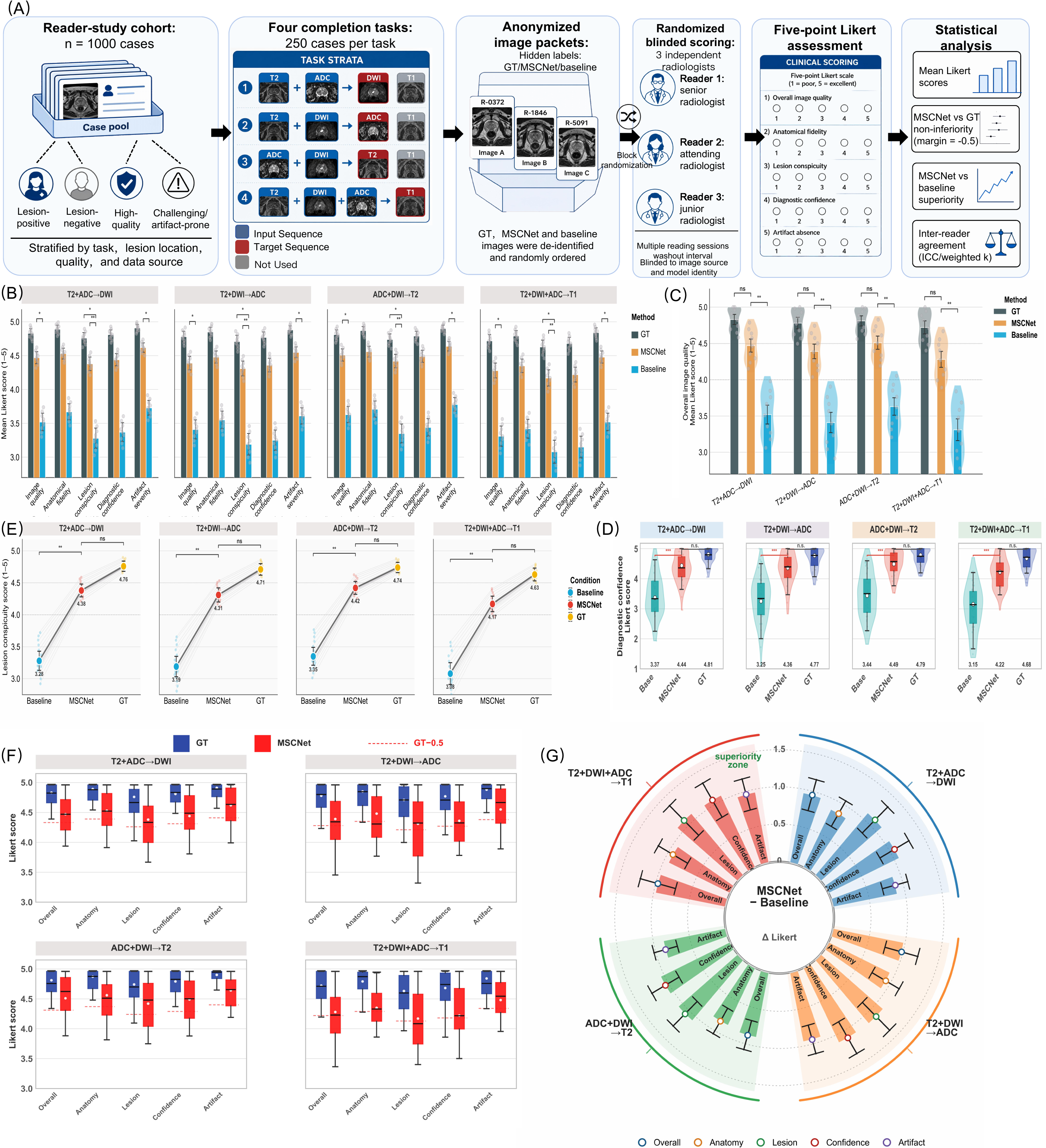}
\caption{\textbf{Blinded radiologist reader study.}
\textbf{A,} Sampling, task strata, anonymization, block randomization, reader experience, Likert endpoints and statistical plan.
\textbf{B,} Mean scores for five clinical dimensions across four completion tasks.
\textbf{C,} Overall image-quality distributions.
\textbf{D,} Diagnostic-confidence distributions.
\textbf{E,} Task-level lesion-conspicuity scores.
\textbf{F,} Reference and MSCNet score distributions relative to the prespecified reference-minus-0.5 margin.
\textbf{G,} Improvement of MSCNet over the baseline across tasks and clinical dimensions.}
\label{fig:readers}
\end{figure}

A blinded reader study assessed 1,000 held-out cases from four representative tasks, with 250 cases per task (Fig.~\ref{fig:readers}A). Acquired reference, MSCNet and baseline images were de-identified, block-randomized and read in multiple sessions by three radiologists. The five prespecified dimensions were overall image quality, anatomical fidelity, lesion conspicuity, diagnostic confidence and absence of artefact.

MSCNet consistently narrowed the gap to acquired images while maintaining a large separation from the baseline (Fig.~\ref{fig:readers}B--E). Task-level overall image-quality scores for baseline, MSCNet and reference images were 3.28, 4.38 and 4.76 for DWI completion; 3.19, 4.31 and 4.71 for ADC completion; 3.35, 4.42 and 4.74 for T2W completion; and 3.08, 4.17 and 4.63 for T1W completion. Diagnostic-confidence scores showed the same ordering: 3.37, 4.44 and 4.81; 3.26, 4.36 and 4.77; 3.44, 4.49 and 4.79; and 3.15, 4.22 and 4.68. The larger residual gap in T1W completion and lesion conspicuity was consistent with the earlier anatomical analysis rather than hidden by the aggregate score.

Because each task contributed 250 cases, the cross-task arithmetic means were 4.71 for reference images, 4.32 for MSCNet and 3.23 for the strongest baseline for overall image quality, and 4.76, 4.38 and 3.31, respectively, for diagnostic confidence (Extended Data Table~\ref{ext:clinical}). Secondary descriptive means were 4.67 versus 4.29 for anatomical fidelity and 4.46 versus 4.07 for lesion conspicuity in lesion-containing cases. Reader- and case-clustered confidence intervals supported the prespecified $-0.5$-point non-inferiority criterion for overall image quality in DWI (difference $-0.38$, 95\% CI $-0.46$ to $-0.30$), ADC ($-0.40$, $-0.48$ to $-0.32$) and T2W completion ($-0.32$, $-0.40$ to $-0.24$), but not T1W completion ($-0.46$, $-0.55$ to $-0.37$). Diagnostic confidence showed the same sequence-specific pattern. Inter-reader agreement for overall image quality was high (ICC 0.87; weighted $\kappa=0.83$). Primary and secondary reader analyses are reported in \SupTable{18} and \SupTable{19}.

\FloatBarrier

\subsection*{Diagnostic validation and selective prediction defined the usable operating range}

\begin{figure}[!htbp]
\centering
\includegraphics[width=\textwidth,height=0.72\textheight,keepaspectratio]{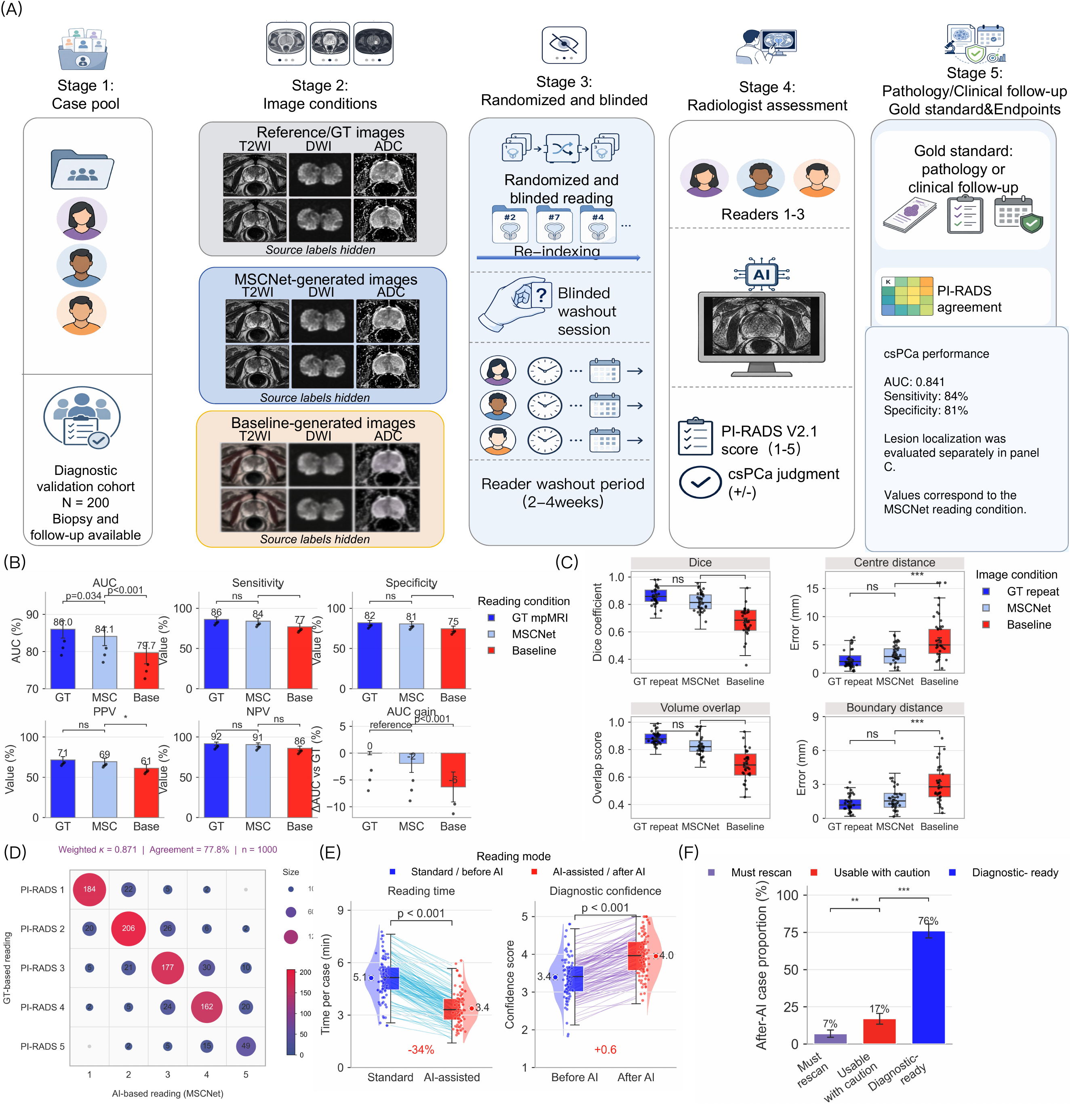}
\caption{\textbf{Diagnostic validation and clinical usability.}
\textbf{A,} Separate diagnostic cohort, image conditions, blinded reading and pathology or clinical follow-up reference standard. Summary values in the endpoint panel correspond to the MSCNet condition and match the quantitative results in B.
\textbf{B,} AUC, sensitivity, specificity, positive predictive value, negative predictive value and AUC difference for acquired, MSCNet and baseline images.
\textbf{C,} Lesion Dice, centre-distance error, volume overlap and boundary-distance error.
\textbf{D,} PI-RADS agreement matrix in the 1,000-case reader-study set.
\textbf{E,} Paired reading time and diagnostic confidence before and after AI assistance.
\textbf{F,} Retrospective usability categories after AI assistance.}
\label{fig:diagnostic}
\end{figure}

A separate 200-case assessment tested whether image completion preserved diagnostic discrimination (Fig.~\ref{fig:diagnostic}A). AUCs for \csPCa\ were 0.860 with acquired images, 0.841 with MSCNet images and 0.797 with baseline-generated images (Fig.~\ref{fig:diagnostic}B). The MSCNet-minus-reference difference was $-0.019$ (95\% CI, $-0.048$ to $+0.010$), compared with $-0.063$ ($-0.094$ to $-0.032$) for the strongest baseline. Sensitivity was 86\%, 84\% and 77\%; specificity was 82\%, 81\% and 75\%; positive predictive value was 71\%, 69\% and 61\%; and negative predictive value was 92\%, 91\% and 86\%. Calibration remained closer to the acquired condition for MSCNet (ECE 0.038, slope 0.94 and Brier score 0.151) than for the baseline (0.067, 0.82 and 0.184) \SupTable{20}. The paired interval is consistent with preservation of diagnostic discrimination under an exploratory $-0.05$ boundary; because that boundary is not used as a confirmatory prespecified and powered margin here, the result is reported as supportive rather than formal diagnostic non-inferiority.

PI-RADS agreement was evaluated across the 1,000-case reader-study set rather than the 200-case pathology subset. Exact agreement between reference-based and MSCNet-based readings was 77.8\%, agreement within one category was 95.6\%, and weighted $\kappa$ was 0.871 (Fig.~\ref{fig:diagnostic}D). In the retrospective workflow analysis, median reading time decreased from 5.1 to 3.4 min per case, diagnostic confidence increased from 3.4 to 4.0, and 76\% of cases were classified as diagnostic-ready, 17\% as usable with caution and 7\% as requiring repeat imaging (Fig.~\ref{fig:diagnostic}E,F and Extended Data Table~\ref{ext:clinical}). The observed workflow benefit was therefore evaluated as an adjunctive use case in which generated images remain linked to the acquired examination. Complete diagnostic and retrospective workflow endpoints are reported in \SupTable{5}.

The locked model was then evaluated in a 186-case external cohort comprising site A, \ChanghaiHospital\ (72 cases; 39\%); site B, \ShanghaiSecondHospital\ (64 cases; 34\%); and site C, \HangzhouBayHospital\ (50 cases; 27\%). Site-level SSIM was 0.798, 0.788 and 0.784, respectively, and diagnostic AUC was 0.831, 0.819 and 0.814. The pooled estimates were SSIM 0.791, LPIPS 0.124, MAE 0.061, AUC 0.823 and ECE 0.042 \SupTable{16}. A method-by-site interaction was not detected ($P=0.55$), although precision was limited by the three external sites \SupTable{24}. The native internal risk threshold ($\tau=0.50$) was transferred without hospital-specific recalibration and retained 37 of 72, 31 of 64 and 23 of 50 cases, respectively; pooled coverage was 48.9\% (91 of 186), with failure detection 0.75, false rejection 0.16, retained-case AUC 0.86 and ECE 0.036 \SupTable{25}. The three-hospital results support transport of the locked model and threshold while retaining site-level calibration and monitoring as part of deployment.

The remaining question was whether unreliable outputs could be recognized before interpretation. We therefore evaluated the operational reconstruction-risk score generated by the study quality-control pipeline. The score was fitted using development and validation data only; no held-out target or radiologist label was used for test-set adaptation. Regional summaries covered the global image, prostate boundary, peripheral zone and transition zone, whereas lesion-region summaries were used only in annotated retrospective analyses. The score correlated with MAE ($\rho=0.58$), LPIPS ($\rho=0.61$), SSIM degradation ($\rho=0.41$) and lesion error ($\rho=0.42$) \SupFig{6}. Median scores increased from 0.16 in radiologist-rated reliable cases to 0.43 in cases usable with caution and 0.79 in unreliable cases ($P<0.001$). Failure-detection AUROC was 0.82 for motion artefact, 0.92 for sequence mismatch, 0.79 for boundary blur and 0.90 for lesion reconstruction error. At the validation-selected threshold of 0.5, failure-detection sensitivity was 0.74, false rejection was 0.18, rejected-case precision was 0.78 and F1 was 0.76. The risk--coverage analysis showed that observed selective risk, defined as one minus diagnostic usability among retained cases, was 0.04 at 25\% coverage, 0.08 at 50\% coverage, 0.14 at 75\% coverage and 0.21 without gating. Thus, the balanced policy achieved diagnostic usability of 0.92 while retaining 50\% of outputs and was not interpreted as reliability at full clinical coverage. The full internal risk--coverage curve and operating points are reported in \SupFig{6} and \SupTable{8}, and the fixed-threshold external transport results in \SupTable{25}. No hospital-specific recalibration, prospective decision-curve analysis or clinical net-benefit analysis was performed.

\FloatBarrier

\section*{Discussion}

This study treats incomplete prostate mpMRI as a problem of information routing, clinical fidelity and selective use rather than image translation alone. The principal task-specific MSCNet models retained the highest SSIM after the comparison set was expanded to unified synthesis, matched-capacity Transformers, modern diffusion models, loss-matched convolutional controls, retrieval and physics-aware baselines. The smaller but persistent advantage under these stronger controls is more informative than the larger gap to legacy baselines: it shows that capacity and optimization explain part, but not all, of the result.

The task hierarchy also constrains the mechanistic and clinical claims. DWI and ADC share diffusion-derived information and should not be treated as equivalent evidence to reconstruction of T2W anatomy. Separating tasks showed mean SSIM advantages of 0.018 in the physically coupled group, 0.024 in cross-contrast T2W synthesis and 0.020 for T1W completion. No universal change in SSIM can be assumed to represent a clinically meaningful difference, and the 0.020 ten-task gain is therefore interpreted as a technical endpoint rather than a diagnostic surrogate. Its clinical-anatomical relevance is supported by the matched-control subset, in which the advantage was larger for lesion fidelity (+0.05) and boundary SSIM (+0.055), together with independent reader, safety and diagnostic analyses. This triangulation supports the interaction between sequence-specific routing, global anatomical context and boundary supervision rather than a claim based on parameter count or global SSIM alone.

The shared-model experiments define both an opportunity and a boundary. One 158.3-million-parameter model reproduced all ten tasks with a mean SSIM reduction of 0.015 relative to the optimized task-specific models, and unified pretraining followed by task-specific fine-tuning recovered the optimized task-specific mean. Zero-shot unseen-combination performance remained lower, however. The evidence therefore supports a clinically evaluated task-specific framework with a unified routing extension; zero-shot performance remained outside the principal claim.

Clinical validation progressed from controlled to less curated settings. Clinically derived artefact transfer isolated five corruption classes, while 204 same-patient repeat acquisitions tested genuine degraded scans against a clean repeat reference. The repeat-scan improvement from SSIM 0.714 to 0.883 supports real-acquisition recovery, but repeated examinations can differ in positioning, physiology and interval change and remain a retrospective reference rather than a perfect ground truth. Similarly, the pooled 186-case external cohort (site A, 72 cases; site B, 64; site C, 50) showed only a modest completion shift and retained an AUC of 0.823. Changhai Hospital and Shanghai Second People's Hospital were institutionally independent of the development centre, whereas Ningbo Hangzhou Bay Hospital is a geographically distinct hospital affiliated with Ren Ji Hospital; the analysis therefore supports multicentre transportability but not validation across three fully independent health systems.

Reader and diagnostic analyses exposed sequence-specific limits that aggregate means would hide. Overall image-quality non-inferiority was supported for DWI, ADC and T2W, but not T1W, and diagnostic confidence showed the same pattern. The 200-case AUC difference between MSCNet and acquired images was small, with a confidence interval that remained above a supportive $-0.05$ boundary; formal diagnostic non-inferiority nevertheless depends on whether that boundary and its power analysis were prespecified before unblinding. These results support preservation of diagnostic information within the evaluated workflow while keeping the acquired examination as the clinical reference.

Selective prediction provides a practical response to the remaining failures, but reliability cannot be separated from coverage. Internally, diagnostic usability increased from 0.79 at full coverage to 0.92 at the validation-selected 50\% operating point. When the native threshold $\tau=0.50$ was transferred externally, it retained 48.9\% of cases, with failure detection 0.75, false rejection 0.16, retained-case AUC 0.86 and ECE 0.036. This site-level degradation shows why threshold transport must be reported explicitly and monitored by hospital. Generated images should remain labelled, displayed beside acquired sequences and withheld when predicted risk is high, anatomy is incomplete or the inputs are severely corrupted.

Limitations remain. The pooled external cohort combined three hospitals, and hospital-specific acquisition distributions, patient flow and calibration may influence the aggregate estimate. Repeat-scan analyses require patient-clustered inference and careful accounting for treatment or interval change. Generated ADC maps should not be used for quantitative ADC measurement without separate calibration, missing anatomical coverage cannot be recreated from absent information, and T1W completion does not recover dynamic contrast kinetics. Prospective multicentre workflow studies, locked health-economic endpoints and post-deployment calibration monitoring remain necessary. Within these boundaries, MSCNet supports a reliability-gated approach to cross-modal completion and enhancement of prostate MRI.

\section*{Online Methods}

\subsection*{Cohorts, governance and task definition}

The study used public and hospital-based prostate MRI data. PI-CAI contained T2W, ADC and high-b-value DWI; the vendor-specific PROSTATEx analysis contained T2W, DWI and ADC.\cite{9,10} The institutional three-sequence and four-sequence cohorts, clinical artefact cohort, reader study and internal diagnostic cohort were obtained at \RenjiHospital, Shanghai, China. The clinical artefact cohort contained 182 unique patients and 262,306 DICOM slices. A locked external cohort contained 186 cases: site A, 72 cases (39\%) from \ChanghaiHospital, Shanghai, China; site B, 64 cases (34\%) from \ShanghaiSecondHospital, Shanghai, China; and site C, 50 cases (27\%) from \HangzhouBayHospital, Ningbo, Zhejiang, China. All external cases had paired completion targets and pathology-confirmed outcomes.

All development splits were performed at the patient level. PI-CAI was divided into 1,000 development, 200 validation and 276 held-out cases; the institutional three-sequence cohort into 2,232, 318 and 639; and the four-sequence cohort into 1,111, 157 and 327. The external cohort, reader study, diagnostic cohort and repeat-scan analyses were not used to select principal model weights. Hospital data were processed retrospectively after de-identification under institutional ethics approval and data-governance agreements. The requirement for written informed consent was waived for retrospective secondary analysis under the applicable local procedures. The verified Ren Ji approval identifier and hospital-level governance roles are reported in the Ethics statement and \SupTable{26}.

For three-sequence data, the targets were DWI from T2W and ADC; ADC from T2W and DWI; and T2W from ADC and DWI. For four-sequence data, the targets were T1W from T2W, DWI and ADC; ADC from T2W, DWI and T1W; DWI from T2W, ADC and T1W; and T2W from DWI, ADC and T1W. The principal model was trained independently for each target configuration. A secondary MSCNet-Shared experiment used one model for all targets, with an available-modality mask, target-modality token and training-time modality dropout. Cohort governance, public-dataset overlap, shared-model data use and exclusion rules are detailed in \SupNote{1} and \SupNote{10}.

\subsection*{Preprocessing, matching and quality control}

DICOM series were identified from metadata and verified using rule-based and manual checks. Images were standardized for orientation, resampled to a common analysis grid and aligned within each examination. Sequence-specific intensity normalization was applied after spatial processing. The prostate region and adjacent context were retained for model input. Cases with corrupted files, incomplete metadata, severe non-correctable misregistration, unreliable lesion annotation or non-diagnostic image content were excluded.

Sequence correspondence was checked at the patient, examination, slice and anatomical-region levels. Rigid or affine alignment was followed by deformable correction when required. Quality was measured using landmark displacement, normalized mutual information, prostate-region Dice and centre displacement. The correction and exclusion analyses are shown in \SupFig{4}; complete rules are provided in \SupNote{2}.

\subsection*{MSCNet}

At scale $s$, modality-specific encoders produced feature maps $F^{(s)}_m$ for each available modality $m$. Each encoder used residual blocks\cite{17} with squeeze-and-excitation recalibration\cite{23} at five scales with 32, 64, 128, 256 and 512 channels. The scale-specific feature tensor was
\[
C^{(s)}=\operatorname{Concat}\left(F^{(s)}_1,\ldots,F^{(s)}_M\right).
\]
A gating network generated modality- and location-dependent weights
\[
g^{(s)}_m=\sigma\!\left(\phi^{(s)}_m(C^{(s)})\right),
\]
and the fused representation was
\[
\widetilde F^{(s)}=\psi^{(s)}\!\left(\operatorname{Concat}\left(g^{(s)}_1\odot F^{(s)}_1,\ldots,g^{(s)}_M\odot F^{(s)}_M\right)\right).
\]
Here, $\sigma$ denotes the sigmoid function, $\odot$ element-wise multiplication and $\psi^{(s)}$ the merge transform. This formulation allowed the contribution of each sequence to vary by feature scale and spatial location.

The deepest fused representation was tokenized and processed by a multi-head self-attention bottleneck.\cite{21} Decoder features were upsampled and combined with fused encoder features through attention gates.\cite{22} Auxiliary predictions at intermediate scales provided deep supervision.\cite{24} An edge branch compared gradients of the prediction and acquired target. Further implementation details are given in \SupNote{3}.

For MSCNet-Shared, absent sequences were represented by zero-filled tensors accompanied by a binary availability mask, and a learned target token specified the requested output. Training sampled valid available-target combinations and applied modality dropout without exposing held-out test cases. The shared model used 158.3 million parameters. A unified pretraining analysis used 3,750 development cases (1,200 PI-CAI development/validation cases and 2,550 institutional three-sequence development/validation cases), followed by task-specific fine-tuning, few-shot adaptation or frozen-encoder transfer. These experiments were secondary and did not replace the task-specific principal models (\SupTable{12} and \SupTable{13}).

\subsection*{Optimization and comparison models}

The full objective was
\[
\mathcal{L}=\mathcal{L}_{\mathrm{char}}
+0.15\mathcal{L}_{\mathrm{SSIM}}
+0.10\mathcal{L}_{\mathrm{MS\text{-}SSIM}}
+0.02\mathcal{L}_{\mathrm{perc}}
+0.05\mathcal{L}_{\mathrm{edge}}
+0.10\mathcal{L}_{\mathrm{wave}}
+0.05\mathcal{L}_{\mathrm{freq}}
+\lambda_{\mathrm{ds}}\mathcal{L}_{\mathrm{deep}}.
\]
The reconstruction terms follow established image-restoration objectives,\cite{25,26,27,45,46} and $\lambda_{\mathrm{ds}}$ denotes the normalized contribution of the auxiliary decoder outputs. Task-specific MSCNet contained 154.1 million parameters and was trained for 200 epochs with AdamW,\cite{44} an initial learning rate of $2\times10^{-4}$, batch size 2, cosine annealing and 10 warm-up epochs.

The legacy comparison set contained DynUNet implemented in MONAI,\cite{20} Pix2Pix,\cite{19} a residual generator,\cite{17} an LSGAN model\cite{39} with a PatchGAN discriminator and spectral normalization,\cite{40} and a conditional diffusion model based on 1,000 diffusion steps and 50-step DDIM sampling.\cite{41,42,43} The expanded comparison set added a unified missing-modality model, a multi-contrast Transformer, modality-masked diffusion, a structure-aware latent-diffusion model, a parameter-matched Transformer, wide DynUNet, DynUNet trained with the MSCNet objective, conditional flow matching, nearest-neighbour retrieval and target-specific signal-model or protocol-regression controls. Data splits, target definitions and evaluation masks were identical. Hyperparameters were selected on validation data with a matched search budget. Complete implementation and fairness details are reported in \SupTable{9}--\SupTable{11} and \SupNote{10}.

\subsection*{Image, spatial and lesion-level evaluation}

Image fidelity was measured using PSNR, SSIM,\cite{25} MS-SSIM,\cite{26} LPIPS,\cite{27} FID\cite{28} and MAE. Spatial evaluation partitioned error into gland, boundary-ring and peri-gland regions and stratified results by apex, mid-gland and base. Volumetric consistency included adjacent-slice similarity, slice-to-slice variation, volumetric smoothness and z-trend correlation. Boundary preservation included edge preservation, boundary SSIM, gradient similarity and capsule sharpness.

Lesion fidelity was assessed using lesion-to-background contrast-to-noise ratio, lesion-region signal-to-noise ratio, contrast ratio, line-profile agreement and radiologist conspicuity. Lesion localization used Dice overlap,\cite{29} centre distance, volume overlap and boundary distance. Evaluation definitions and aggregation rules are given in \SupNote{5}.

\subsection*{Artefact transfer and same-patient repeat-scan evaluation}

Clinically observed artefact patterns were extracted from low-quality acquisitions and categorized as susceptibility, motion ghosting, banding, Gibbs ringing or chemical shift. Candidate components were screened for topology and texture, aligned with clean targets and transferred in image or frequency space. Pairing required anatomical correspondence, lesion-boundary preservation and SSIM of approximately 0.8 between the clean target and controlled degraded image. Restoration was evaluated against the quality-controlled target with six image-quality metrics and radiologist safety review.

A distinct repeat-scan analysis used 204 degraded--repeat pairs from 182 unique patients. One patient could contribute more than one adjudicated stratum. The clean reference was a same-patient repeat acquisition obtained within four weeks; treatment status, interval, scanner and protocol correspondence were recorded. Categories were motion blur, susceptibility distortion, low-SNR high-b-value DWI, geometric distortion, zipper artefact, ghosting and radiofrequency inhomogeneity. Statistical inference used patient-clustered paired comparisons. The controlled-transfer and repeat-scan taxonomies were analysed separately (\SupTable{3}, \SupTable{17} and \SupNote{11}).

\subsection*{Reader study, diagnostic assessment and external evaluation}

The reader study sampled 1,000 held-out cases across four tasks, with 250 cases per task. Acquired, MSCNet and baseline images were de-identified and block-randomized. Three radiologists read the images in sessions with washout and scored overall quality, anatomical fidelity, lesion conspicuity, diagnostic confidence and artefact absence on five-point scales. The acquired-reference comparison used a prespecified $-0.5$-point margin. Primary task-level differences and 95\% confidence intervals were estimated with models retaining reader and case clustering; non-inferiority was supported only when the lower confidence bound exceeded the margin. Agreement was summarized by ICC and weighted $\kappa$.\cite{31,32}

The diagnostic assessment used 200 cases with biopsy or clinical follow-up. Readers assigned PI-RADS version 2.1 categories and binary \csPCa\ judgements. AUCs were compared using paired methods,\cite{30} and calibration was summarized by expected calibration error, calibration slope and Brier score. The $-0.05$ AUC boundary was used only as an exploratory supportive benchmark and was not treated as a confirmatory non-inferiority margin. PI-RADS agreement was calculated separately in the 1,000-case reader cohort.

The external cohort contained 186 cases across \ChanghaiHospital\ (site A; 72 cases; 39\%), \ShanghaiSecondHospital\ (site B; 64 cases; 34\%) and \HangzhouBayHospital\ (site C; 50 cases; 27\%), with paired completion targets and pathology-confirmed outcomes. Site A used a 3.0-T Siemens MAGNETOM Prisma system, site B a 3.0-T Philips Ingenia system and site C a 1.5-T GE SIGNA Artist system; protocol differences included T2W resolution, slice thickness and DWI $b$-values \SupTable{23}. The trained reconstruction models, preprocessing rules and internal risk threshold were fixed before external evaluation. Completion and diagnostic endpoints were reported separately even when the same case contributed to both. No hospital-specific training, fine-tuning or threshold recalibration was performed. Site-level case spectrum, performance, interaction and threshold-transport results are provided in \SupTable{16} and \SupTable{22}--\SupTable{25}.

\subsection*{Safety adjudication, uncertainty and rejection}

Safety events were defined as suspected false lesion, reduced lesion conspicuity, blurred lesion margin, partial lesion erasure, non-diagnostic output or input-output inconsistency. Events were adjudicated by the radiologists and assigned to acquisition-dominant, lesion-dominant, registration/model-dominant, mixed or unassigned contributors.

The quality-control analysis used a normalized voxel-wise reconstruction-risk map, denoted $U(x)$, generated by the implementation archived with the review code. Model fitting and operating-point selection used development and validation data only, and the risk score was fixed before evaluation on held-out internal and external cohorts. The score was interpreted operationally as reconstruction risk rather than as a decomposition of aleatoric and epistemic uncertainty.

The primary case-level score was the maximum of the 95th-percentile values within the global image, prostate boundary, peripheral zone and transition zone. Lesion-region summaries were calculated only retrospectively in cases with lesion annotations and were not required for the primary score. Failure detection was evaluated by AUROC for motion artefact, sequence mismatch, boundary blur and lesion reconstruction error. Thresholds from 0.2 to 0.7 were assessed using failure-detection sensitivity, false-rejection rate, rejected-case precision and F1. Coverage was the proportion of outputs retained below a threshold. Selective risk was the mean failure indicator among retained cases; for the clinical-usability analysis it was additionally reported as one minus diagnostic usability. Risk--coverage curves were generated by sweeping the fixed score threshold. Expected calibration error was computed from fixed-width score bins. The archived code and configuration files specify the estimator implementation used for the reported experiments.

The threshold of 0.5 and the conservative, balanced and permissive policies were selected on validation data and evaluated internally without test-set retuning. The locked internal threshold was then transferred directly to the 186-case external cohort without hospital-specific recalibration. Coverage, retained count, failure detection, false rejection, retained-case AUC and calibration were reported jointly for external threshold transport. No decision-curve, cost-effectiveness or prospective clinical net-benefit analysis was performed. Exact definitions and external transport results are provided in \SupNote{9}, \SupNote{13}, \SupTable{8}, \SupTable{21} and \SupTable{25}.

\subsection*{Statistical analysis and reporting}

Continuous values were summarized by mean and standard error or median and interquartile range, as appropriate. Paired normally distributed differences used paired $t$-tests; non-normal paired or ordinal outcomes used Wilcoxon signed-rank tests. Patient clustering was retained for multiple repeat acquisitions. Correlations were summarized using Spearman's $\rho$. AUC comparisons used paired DeLong or paired bootstrap procedures.\cite{30} Weighted $\kappa$\cite{31} and ICC\cite{32} quantified reader agreement. Reader non-inferiority models retained crossed reader and case effects, and conclusions were based on confidence-interval lower bounds rather than point estimates alone. Multiple comparisons were controlled using the Benjamini--Hochberg procedure.\cite{33}

Thresholds were chosen on validation data; held-out internal and external risk--coverage results were reported with corresponding coverage and without test-set retuning. External threshold transport was assessed using the native internal threshold. All model and task summaries were calculated at case level, and task groups were prespecified as physically coupled DWI--ADC, cross-contrast T2W and auxiliary T1W. No formal health-economic, cost-effectiveness or decision-curve analysis was conducted. Reporting was guided by CLAIM, STARD and STARD-AI.\cite{34,35,36}

\section*{Ethics approval and consent}

The retrospective institutional study was approved by the Ethics Committee of Ren Ji Hospital, Shanghai Jiao Tong University School of Medicine (approval no. \RenjiIRB) and was conducted in accordance with the Declaration of Helsinki and applicable data-protection requirements. The requirement for written informed consent for retrospective use of de-identified institutional data was waived by the approving committee. De-identified external data from \ChanghaiHospital, \ShanghaiSecondHospital\ and \HangzhouBayHospital\ were analysed under the applicable local ethics, consent-waiver and data-governance procedures and institutional data-use agreements. Public PI-CAI and PROSTATEx data were used under their respective access and governance conditions.

\section*{Data availability}

PI-CAI and PROSTATEx are available from their public repositories subject to the applicable access terms. Institutional imaging and clinical data are not publicly released because they contain protected health information. Requests for access to de-identified institutional data should be directed to the co-corresponding authors and will be reviewed by the relevant institutional data-access committee. Access requires an approved research purpose, applicable ethics approval, a data-use agreement and compliance with relevant privacy and data-protection requirements.

\section*{Code availability}

Code, configuration files and inference instructions for the reported analyses are available to editors and reviewers at \url{\CodeReviewURL}. A versioned public repository and archival DOI are planned for the published record.

\section*{Funding}

This work was supported by the National Natural Science Foundation of China (No. 82371912) and the Science and Technology Commission of Shanghai Municipality (No. YDZX20243100003001).

\section*{Acknowledgements}

The authors thank the participating radiologists and the clinical data-management teams at Ren Ji Hospital, Changhai Hospital, Shanghai Second People's Hospital and Ningbo Hangzhou Bay Hospital for support with blinded review, data curation and governance.

\section*{Author contributions}

All authors made substantial contributions to one or more aspects of study conception, methodology, investigation, data curation, formal analysis, software, validation, visualization or manuscript preparation. All authors reviewed and approved the final manuscript. Yang Liu, Dacheng Tao and Guangyu Wu are co-corresponding authors.

\section*{Competing interests}

The authors declare no competing interests.

\clearpage
\FloatBarrier

\clearpage
\section*{Extended Data}
\setcounter{table}{0}
\renewcommand{\tablename}{Extended Data Table}
\begin{table}[H]
\centering
\caption{\textbf{Study cohorts and analytical roles.} Splits were defined at the patient level. PROSTATEx was treated as a vendor-specific public cohort rather than independent external validation.}
\label{ext:cohorts}
\small
\begin{tabularx}{\textwidth}{@{}P{0.17\textwidth}P{0.23\textwidth}P{0.17\textwidth}P{0.13\textwidth}Y@{}}
\toprule
Cohort & Source & Sequences & Size or split & Primary role\\
\midrule
PI-CAI & Public, multicentre & T2W, ADC, high-b-value DWI & 1,476; 1,000/200/276 & Public-domain development and held-out evaluation of 2-to-1 tasks\\
PROSTATEx & Public, Siemens & T2W, DWI, ADC & 346 & Vendor-specific public-cohort analysis\\
Institutional three-sequence & Ren Ji Hospital, Shanghai, China & T2W, DWI, ADC & 3,189; 2,232/318/639 & Training, validation and testing of 2-to-1 tasks\\
Institutional four-sequence & Ren Ji Hospital, Shanghai, China & T2W, DWI, ADC, T1W & 1,595; 1,111/157/327 & Training, validation and testing of 3-to-1 tasks\\
Clinical artefact & Ren Ji Hospital archive & Degraded prostate MRI sequences & 182 patients; 262,306 DICOM slices; 204 repeat pairs & Controlled artefact transfer, same-patient repeat-scan validation and safety evaluation\\
Reader study & Ren Ji Hospital held-out cohort & Four representative tasks & 1,000; 250 per task & Three-reader blinded assessment\\
Diagnostic assessment & Ren Ji Hospital separate clinical set & Reference, MSCNet and baseline conditions & 200 & csPCa discrimination and workflow endpoints\\
Three-hospital external cohort & Site A: Changhai Hospital (72; 39\%); site B: Shanghai Second People's Hospital (64; 34\%); site C: Ningbo Hangzhou Bay Hospital (50; 27\%) & T2W, DWI, ADC & 186 total & Locked pooled completion, diagnostic and threshold-transport evaluation\\
\bottomrule
\end{tabularx}
\end{table}

\begin{table}[H]
\centering
\caption{\textbf{Main quantitative performance of MSCNet.} Values in parentheses are absolute changes relative to the strongest non-MSCNet baseline. PSNR and SSIM are higher-is-better; LPIPS and MAE are lower-is-better.}
\label{ext:quantitative}
\scriptsize
\begin{tabularx}{\textwidth}{@{}P{0.11\textwidth}P{0.045\textwidth}Yrrrr@{}}
\toprule
Dataset & Task & Input $\rightarrow$ target & PSNR & SSIM & LPIPS & MAE\\
\midrule
PI-CAI & P1 & T2W + ADC $\rightarrow$ high-b DWI & 25.91 (+0.53) & 0.8600 (+0.0180) & 0.0890 ($-$0.0100) & 0.0384 ($-$0.0036)\\
PI-CAI & P2 & T2W + high-b DWI $\rightarrow$ ADC & 28.43 (+0.48) & 0.8890 (+0.0170) & 0.0700 ($-$0.0110) & 0.0261 ($-$0.0029)\\
PI-CAI & P3 & ADC + high-b DWI $\rightarrow$ T2W & 23.07 (+0.66) & 0.7300 (+0.0240) & 0.1480 ($-$0.0160) & 0.0573 ($-$0.0047)\\
Institutional & A & T2W + ADC $\rightarrow$ DWI & 24.17 (+0.46) & 0.8460 (+0.0180) & 0.0950 ($-$0.0110) & 0.0621 ($-$0.0039)\\
Institutional & B & T2W + DWI $\rightarrow$ ADC & 27.46 (+0.44) & 0.8660 (+0.0180) & 0.0830 ($-$0.0100) & 0.0382 ($-$0.0038)\\
Institutional & C & ADC + DWI $\rightarrow$ T2W & 21.59 (+0.61) & 0.7140 (+0.0240) & 0.1620 ($-$0.0160) & 0.0662 ($-$0.0048)\\
Institutional & D & T2W + DWI + ADC $\rightarrow$ T1W & 25.06 (+0.48) & 0.7960 (+0.0200) & 0.1130 ($-$0.0120) & 0.0378 ($-$0.0032)\\
Institutional & E & T2W + DWI + T1W $\rightarrow$ ADC & 27.82 (+0.46) & 0.8790 (+0.0180) & 0.0810 ($-$0.0100) & 0.0351 ($-$0.0029)\\
Institutional & F & T2W + ADC + T1W $\rightarrow$ DWI & 24.53 (+0.45) & 0.8600 (+0.0190) & 0.0900 ($-$0.0110) & 0.0573 ($-$0.0037)\\
Institutional & G & DWI + ADC + T1W $\rightarrow$ T2W & 22.31 (+0.63) & 0.7380 (+0.0250) & 0.1490 ($-$0.0170) & 0.0617 ($-$0.0053)\\
\bottomrule
\end{tabularx}
\end{table}

\begin{table}[H]
\centering
\caption{\textbf{Clinical reader-study and diagnostic-validation summary.} Reader scores were obtained in 1,000 held-out cases; diagnostic discrimination was evaluated in a separate 200-case set.}
\label{ext:clinical}
\scriptsize
\begin{tabularx}{\textwidth}{@{}P{0.17\textwidth}P{0.23\textwidth}Y P{0.23\textwidth}@{}}
\toprule
Domain & Endpoint & Comparison or statistic & Result\\
\midrule
Reader study & Overall image quality & Reference / MSCNet / strongest baseline & 4.71 / 4.32 / 3.23\\
Reader study & Anatomical fidelity & Reference / MSCNet; secondary descriptive endpoint & 4.67 / 4.29\\
Reader study & Lesion conspicuity & Reference / MSCNet; lesion-containing cases & 4.46 / 4.07\\
Reader study & Diagnostic confidence & Reference / MSCNet / strongest baseline & 4.76 / 4.38 / 3.31\\
Reader study & Artefact absence & Reference / MSCNet / strongest baseline & 4.82 / 4.56 / 3.41; $P<0.001$\\
Reader agreement & Inter-reader agreement & ICC; weighted $\kappa$ & 0.87; 0.83\\
Reader non-inferiority & Overall quality & DWI / ADC / T2W / T1W & Supported / supported / supported / not demonstrated\\

PI-RADS & Exact and within-one agreement & Reference-based versus MSCNet-based reads; $n=1,000$ & 77.8\%; 95.6\%; weighted $\kappa=0.871$\\
Diagnostic validation & csPCa discrimination & Reference / MSCNet / baseline & AUC 0.860 / 0.841 / 0.797\\
Diagnostic validation & AUC difference (95\% CI) & MSCNet-reference / baseline-reference & $-0.019$ [$-0.048$, $+0.010$] / $-0.063$ [$-0.094$, $-0.032$]\\
External validation & Completion / diagnostic & Three hospitals; 186 cases (A/B/C: 72/64/50) & SSIM 0.791; LPIPS 0.124; MAE 0.061; AUC 0.823\\

Diagnostic validation & Sensitivity / specificity & Reference / MSCNet / baseline & 86/82\%; 84/81\%; 77/75\%\\
Diagnostic validation & PPV / NPV & Reference / MSCNet / baseline & 71/92\%; 69/91\%; 61/86\%\\
Workflow assessment & Reading time & Before versus after assistance & 5.1 to 3.4 min; $-34\%$\\
Workflow assessment & Diagnostic confidence & Before versus after assistance & 3.4 to 4.0; $+0.6$\\
Workflow assessment & Usability category & Must rescan / caution / diagnostic-ready & 7\% / 17\% / 76\%\\
\bottomrule
\end{tabularx}
\end{table}

\end{document}